# Laser Resonance Ionization Spectroscopy of Thorium

Ruohong Li[1,3*,4], Yuan Liu[2], D. W. Stracener[2], Jens Lassen[1,5,6]

[1]TRIUMF - Canada's Particle Accelerator Centre, Vancouver, BC V6T 2A3, Canada
[2]Physics Division, Oak Ridge National Laboratory, Oak Ridge, TN 37831, USA
[3]Department of Physics, University of Windsor, Windsor, ON N9B 3P4, Canada
[4]Department of Astronomy and Physics, Saint Mary's University, Halifax, NS B3H 3C3, Canada
[5]Department of Physics and Astronomy, University of Manitoba, Winnipeg, MB R3T 2N2, Canada
[6]Department of Physics, Simon Fraser University, Burnaby, BC V5A 1S6, Canada

**Abstract**

High-lying Rydberg and autoionizing (AI) states of thorium (Th) have been studied via resonance laser ionization spectroscopy at both TRIUMF Canada's particle accelerator centre and Oak Ridge National Lab (ORNL). Multiple Rydberg series converging to the ionization potential (IP) were observed via different stepwise laser excitation schemes and were assigned to be the $6d^2 7s$ ($^4F_{3/2}$) $np$, $nd$, and $nf$ series. Analysis of these series enabled the determination of the IP to be 50868.735(54) cm$^{-1}$, which improved the precision by two orders of magnitude over the current adopted NIST value of 50867(2) cm$^{-1}$. Additionally, four AI Rydberg series were identified and assigned to $nf$ and $nd$ series converging to the $6d^2 7s$ $^4F_{5/2}$ and $6d7s^2$ $^2D_{3/2}$ metastable states of Th$^+$. The measured energies of the Rydberg and AI Rydberg states are reported, and observed perturbations within the series are discussed.

## 1. Introduction

Thorium (Th, Z = 90) is an actinide element with four valence electrons (quadrivalent) in its ground-state electronic configuration. The four valence electrons can be present in the orbits of *5f, 6d, 7p*, and *7s*, giving rise to many possible excited configurations. The interactions between the open subshells and different configurations result in a great number of energetically close-lying electronic states [1]. Further, strong spin-orbit interactions and significant relativistic effects [2] in the atomic orbitals are expected. Therefore, the atomic spectra of Th are rich and complex, which makes them challenging for theoretical modeling and simulations. An experimental investigation is essential for providing benchmark data to validate and refine theoretical models, and to advance the understanding of interactions among different orbital electrons in the extended actinide series.

Measurements of Th spectra date back to the 1910s and 1920s [3,4]. The first list of energy levels and the ground state configurations of neutral Th was published as a dissertation in 1946 [5]. To date, tens of thousands spectral lines and over 700 energy levels of Th have been compiled [6,7]. Most of this data came from the emission spectra of traditional light sources such as electrodeless discharge [8,9] and hollow-cathode lamps [10–13]. The level assignments are limited to energies well below the ionization potential (IP). High-lying levels above 49000 cm$^{-1}$ were lacking in the databases, due to their weak transition strengths and low populations excited in traditional emission spectra methods.

* corresponding email: ruohong@triumf.ca

In the 1980s and 1990s, laser resonance ionization spectroscopy (RIS) was employed to investigate the spectra of Th [14–16] and to determine its IP [15,17], as well as for elemental ultra-trace analysis and isotope-ratio measurements [18,19]. More recently, in-source RIS was applied to study the nuclear properties of Th isotopes and isomers [20,21]. Through stepwise resonant photoexcitation, RIS enables access to highly excited states. Leveraging this capability, Raeder et al. [20,22] have observed many high-lying levels near and above the IP of Th.

Nevertheless, Rydberg progressions in Th have been scarcely observed. It is well known that the observation of Rydberg series in most actinides is challenging due to the spectral complexity and strong perturbations caused by dense multivalence states [23,24]. Johnson et al. [15] showed spectral features that might correspond to Rydberg states via two-step RIS, but the resolution was too poor to identify the individual states. Goncharov et al. [25] reported the even-parity Rydberg series $6d^27s\ ns$ and $6d^27s\ nd$, converging to the IP. However, no level energies of those Rydberg members were reported.

IP is an important atomic property of an element that determines its chemical behavior and affects the accuracy of theoretical calculations of the atomic-level energies. The IP of Th has been measured using RIS in the presence of an external electric field [15,17]. Johnson et al. [15] determined the IP to be 50890(20) cm$^{-1}$ from the onset of photoionization. Trautmann [17] improved this value to be 50867(2) cm$^{-1}$ by extrapolating the photoionization threshold to zero field strength based on the saddle-point model. A more accurate determination could come from the convergence limit of a Rydberg series. Goncharov et al. [25] obtained IP = 50868.71(8) cm$^{-1}$ using the $6d^27s\ ns$ Rydberg series, but gave no detailed data on the Rydberg-state energies. In this work, two-step and three-step RIS were used to study odd- and even-parity Rydberg states and autoionizing (AI) Rydberg states to further improve the precision of the IP and study the (AI) Rydberg progressions and the perturbations within.

## 2. Experimental Setup

The experiments were performed at both TRIUMF and Oak Ridge National Lab (ORNL) using two individual apparatuses with different ion sources, ion optics, and laser systems. The experimental setup at ORNL is described in detail in [26–28]. At TRIUMF, the experiment was conducted using the offline laser ion source test stand (LIS-STAND) [29] as shown in Fig. 1. The Th sample was made of 50 μl water-based Th solution (PerkinElmer Inc, 1 μg/μl Th in the form of Th(NO$_3$)$_x$ in 2% HNO$_3$) dried on a small piece of Ti foil in a 110℃ oven. The dried foil was folded and placed inside a Ta crucible of 20 mm length and 3 mm inner diameter. Under 10$^{-6}$ Torr vacuum, the crucible was resistively heated to a temperature of 2000℃. The evaporated Th atoms were stepwise resonantly excited and ionized by pulsed laser light from multiple lasers. The ions were guided through a 50-mm-long radio-frequency quadrupole (RFQ) [30], extracted and accelerated to 10 keV. The ion beam was deflected 90° vertically away from the incident laser beam path and detected by a channel electron multiplier after mass selection via a commercial quadrupole mass spectrometer (QMS, ABB Extrel MAX-300). The QMS was operated with a mass resolution of Δm ~ 1 amu, which was accomplished by deaccelerating the input ion beam to an energy below 50 eV.

The laser system at TRIUMF consisted of two birefringent-filter-tuned (BRF-tuned) and one grating-tuned titanium:sapphire (Ti:Sa) laser. They were simultaneously pumped by a pulsed, frequency-doubled Nd:YAG laser at 10 kHz repetition rate with the total output power of 28 W. The Ti:Sa lasers typically produced a fundamental output power of 1 – 2 W with a tunable wavelength range of 680 – 980 nm. Through intra-cavity frequency doubling, 200 – 500 mW of second-harmonic laser light could be generated. The pulse length of the Ti:Sa lasers was about 50 ns (FWFH). Due to excellent stability and high output power, the BRF-tuned lasers were used for the first and second excitation steps at specific transition wavelengths to populate the desired intermediate levels. The grating-tuned laser was continuously tunable over a broad wavelength range of 690 – 920 nm [31]. Hence, it was utilized to scan for Rydberg and AI transitions from those intermediate levels.

The two-step and three-step excitation schemes of Th are shown in Fig. 2. For the first excitation step (FES), blue laser light at 372.049 nm (all laser/transition wavelengths in this paper are values in vacuum) excited Th atoms from the ground state to the intermediate metastable state at 26878.162 cm$^{-1}$. The blue laser was generated by a BRF-tuned Ti:Sa laser with intra-cavity doubling using a BBO crystal [32]. In the three-step (blue-red-red) excitation schemes, the second excitation step (SES) used NIST-listed known energy levels as the upper states and was driven by another BRF-tuned Ti:Sa laser, with a typical power of ~ 1 W at its fundamental wavelength. The grating-tuned Ti:Sa laser was scanned to obtain the ion signals from Rydberg and AI resonances.

For two-step (blue-blue) excitation schemes, the frequency-doubled output of the grating-tuned Ti:Sa laser required continuous scanning. This was achieved using an external frequency-doubling system capable of auto-tracking the phase matching of an angle-tuned nonlinear crystal for second-harmonic generation (SHG). The SHG crystal was a type-I LBO crystal (5mm × 8mm × 10mm long) cut for 760 nm at an operating temperature of 21 °C. The fundamental output of the grating-tuned laser was focused into the crystal by a $f = 50$ mm plano-convex lens. The spatial walk-off of the frequency-doubled light far off 760 nm was compensated by a commercial beam stabilization system (TEM Messtechnik BeamLock-4D) [33] to ensure laser-beam overlap in the ionization region unchanged throughout the extended laser frequency scans. The typical fundamental laser power of the grating-tuned Ti:Sa laser was 1 – 2 W, and the frequency-doubled output was 100 – 250 mW.

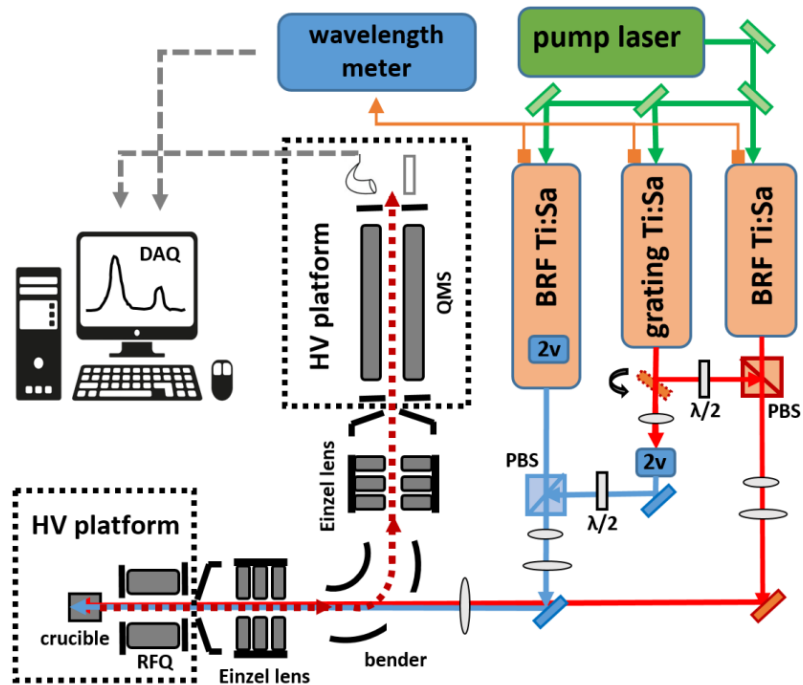

Fig. 1. Schematic of the experimental setup at TRIUMF. BRF: birefringent filter; PBS: polarizing beam splitter cube; λ/2: half waveplate; 2v: nonlinear crystal for frequency doubling; RFQ: radio frequency quadrupole; QMS: quadrupole mass spectrometer; DAQ: data acquisition.

To achieve 3 mm-diameter foci inside the crucible, all the laser beams were expanded 2× – 4× by a Galileo telescope before beam combination via polarizing beam splitters and dichroic mirrors. The combined laser beams were finally focused by an uncoated 2" lens ($f = 5$ m) into the crucible located approximately 5 m away from the laser table. Temporal overlap of the laser pulses was achieved by adjusting the pump laser power distribution between the lasers and by using Pockels cells inside the laser cavities. The laser wavelengths were measured and monitored by a HighFinesse WS/6 wavelength meter that was multiplexed with a 4-channel fiber switcher. The wavelength meter was repeatedly calibrated to a polarization-stabilized HeNe laser with a wavelength accuracy of $10^{-8}$ (Melles Griot 05 STP 901/903). The wavelength systematic uncertainty is 600 MHz (0.02 cm$^{-1}$) according to 3σ criterion.

## 3. Experimental Procedures and Results

The excitation schemes investigated are shown in Fig. 2, where the A schemes represent three-step (blue-red-red) schemes, and the B schemes represent two-step (blue-blue) schemes. These schemes were designed to study energy levels in the vicinity of different Rydberg convergence limits. For example, scheme A1 was aimed at the energy region around the IP of neutral Th (i.e., ground state of Th$^+$), A2 for the 1$^{st}$ metastable state of Th$^+$, and A3 for the 2$^{nd}$ metastable state of Th$^+$. The same denotation applies to the B schemes. ORNL and TRIUMF investigated blue-red-red schemes with three different intermediate levels to study the odd-parity Rydberg states near the IP. TRIUMF did further studies on the odd- and even-parity AI Rydberg states using both blue-red-red and blue-blue excitation schemes.

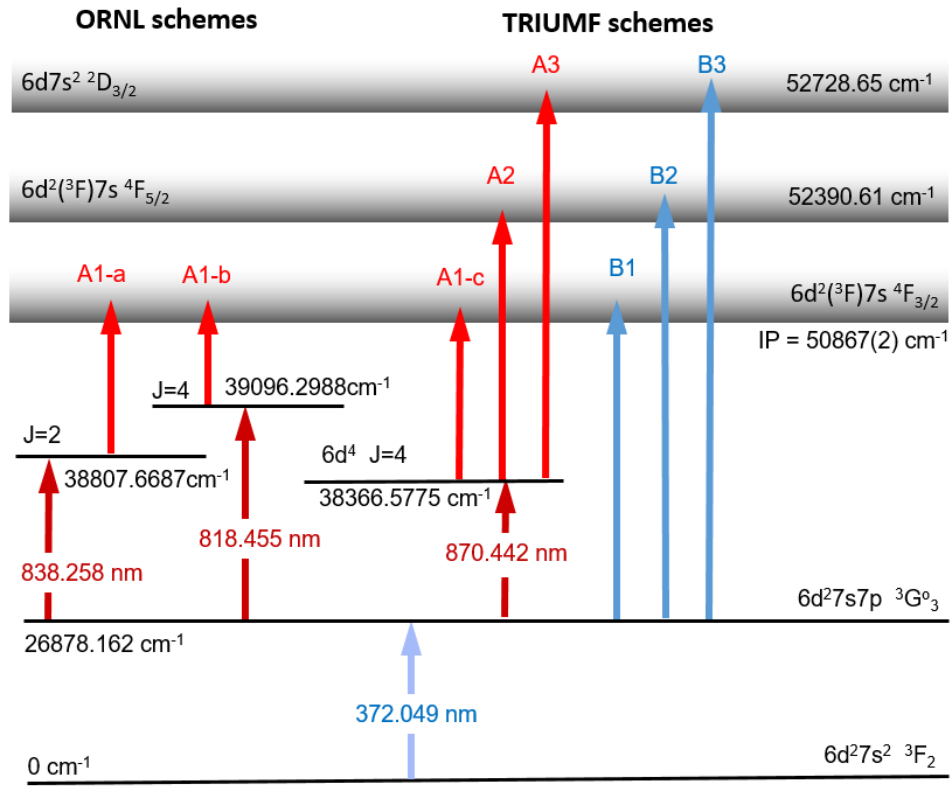

Fig. 2. Laser ionization schemes of Th investigated by ORNL (scheme A1-a, -b) and TRIUMF (scheme A1-c, A2, A3, B1, B2, B3). The energies and configurations of the atomic states, and the IP value are from the NIST ADS database [7] and are presented to their best-known precision before this work. Blue and red arrows refer to the blue/UV-light and infrared-light transitions, respectively.

## 3.1 Rydberg spectra below the IP

Fig. 3 shows the Rydberg and AI spectra obtained using the A1 schemes at ORNL and TRIUMF. With the same FES at 372.049 nm, ORNL used the red transitions at 838.258 nm and 818.455 nm as the SES to the intermediate levels at 38807.6687(9) cm⁻¹ ($J = 2$) and 39096.2986(7) cm⁻¹ ($J = 4$), respectively. TRIUMF used a different SES: the 870.442 nm transition to the level at 38366.5775(8) cm⁻¹ with a known electronic configuration $6d^4$ $J = 4$. The fundamental radiation (690 – 920 nm) of the grating-tuned Ti:Sa lasers was used as the third excitation step (TES) to access the high-lying Rydberg states right below the IP. The Th atoms excited to bound Rydberg states were subsequently ionized in the hot crucible, likely due to absorption of an additional laser photon, blackbody radiation, or thermal collisions within the hot-cavity crucible. The laser power in the laser-atom interaction region was 50 – 300 mW for the FES, and 1 – 2 W for the SES and TES.

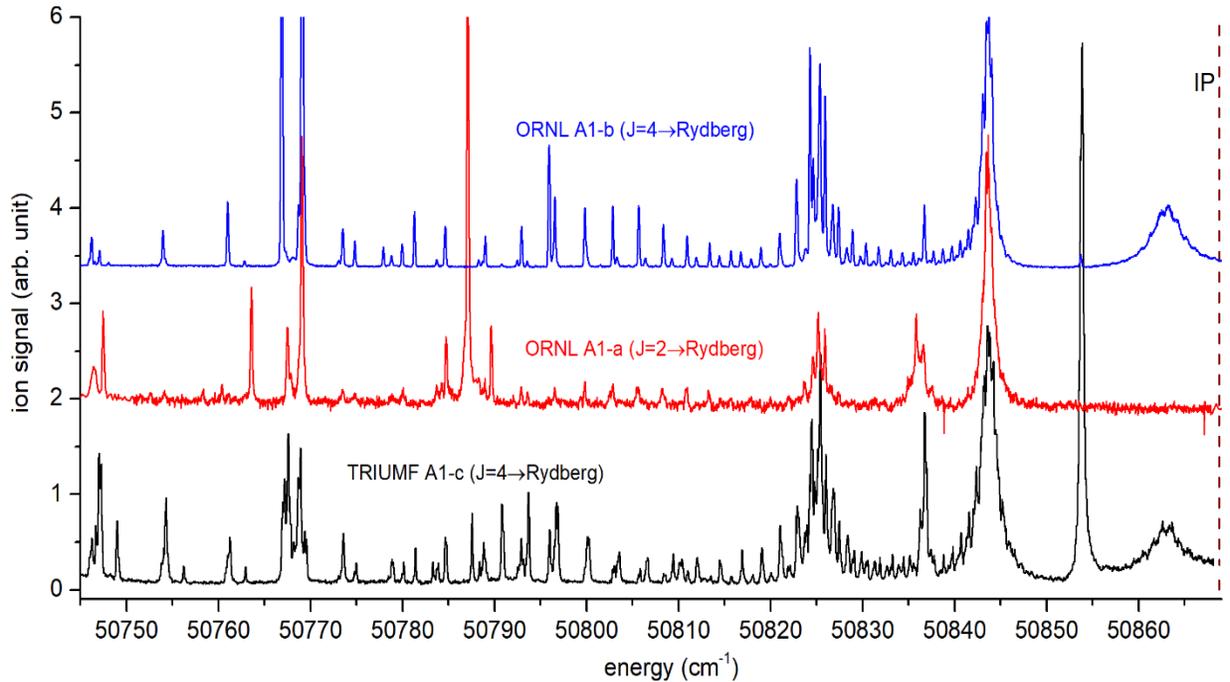

Fig. 3. Resonance ionization spectra observed via schemes A1-a and A1-b at ORNL, and scheme A1-c at TRIUMF.

Table 1. Energies of the intermediate levels measured in this work, compared with the NIST values.

| level | energy (cm⁻¹) | measured by | NIST value (cm⁻¹) |
|---|---|---|---|
| $J = 2$ | 38807.60(6) | ORNL | 38807.6687(9) |
| $J = 4$ | 39096.36(15) | ORNL | 39096.2986(7) |
| $6d^4$ $J = 4$ | 38366.56(15) | TRIUMF | 38366.5775(8) |

The energy values of the second excited levels measured at ORNL and TRIUMF are shown in Table 1. The results are consistent with the values listed in the NIST database [8]. Hence, the NIST values with higher precision were used in this work to calculate the energies of higher levels excited from these SES (Fig. 2). The energies of the identified Rydberg states were extracted by fitting the resonance peaks to a Gaussian profile. Each spectrum was measured multiple times to check reproducibility and to estimate the statistical uncertainty of the resonance centroids. To avoid errors caused by DAQ delays, the spectra were scanned in different scanning directions and speeds. The uncertainty of the extracted Rydberg-state energies was found to be 0.06 cm$^{-1}$, 0.10 cm$^{-1}$, and 0.15 cm$^{-1}$ for the spectra of A1-a, A1-b, and A1-c, respectively, which included the statistical uncertainty of the measurements and systematic uncertainty of the experimental system (see details of error budget in Ref. [28,33]).

Several Rydberg series can be identified in Fig. 3. Being excited from the $J = 2$ and $J = 4$ states with even parity by electric dipole transition, these Rydberg series are expected to be the odd-parity $6d^2 7s$ ($^4F_{3/2}$) $np$ or $nf$ series based on electric-dipole-transition selection rules:

$J = 2 \rightarrow 6d^2 7s$ ($^4F_{3/2}$) $np_{1/2}$, $J = 1, 2$    $J = 4 \rightarrow 6d^2 7s$ ($^4F_{3/2}$) $np_{3/2}$, $J = 3$

  $np_{3/2}$, $J = 1, 2, 3$

$J = 2 \rightarrow 6d^2 7s$ ($^4F_{3/2}$) $nf_{5/2}$, $J = 1, 2, 3$    $J = 4 \rightarrow 6d^2 7s$ ($^4F_{3/2}$) $nf_{5/2}$, $J = 3, 4$

  $nf_{7/2}$, $J = 2, 3$            $nf_{7/2}$, $J = 3, 4, 5$

For the heavy actinides, spin-orbit interactions are stronger than electrostatic interactions [34] and thus $jj$-coupling is more appropriate to describe these Rydberg states. In this work, we used $6d^2 7s$ ($^4F_{3/2}$) as the principal configuration for the Th ionization limit (i.e., ground state of Th$^+$), which is different from $6d7s^2$ ($^2D_{3/2}$) listed in the NIST database but consistent with what listed in the Aimé Cotton Lab database [6], This is based on the leading configuration of $6d^2 7s$ ($^4F_{3/2}$) with a 43% contribution, compared to 27% for $6d7s^2$ ($^2D_{3/2}$). Goncharov et al. [24] also used $6d^2 7s$ ($^4F_{3/2}$) as the principal configuration for the Th ionization limit.

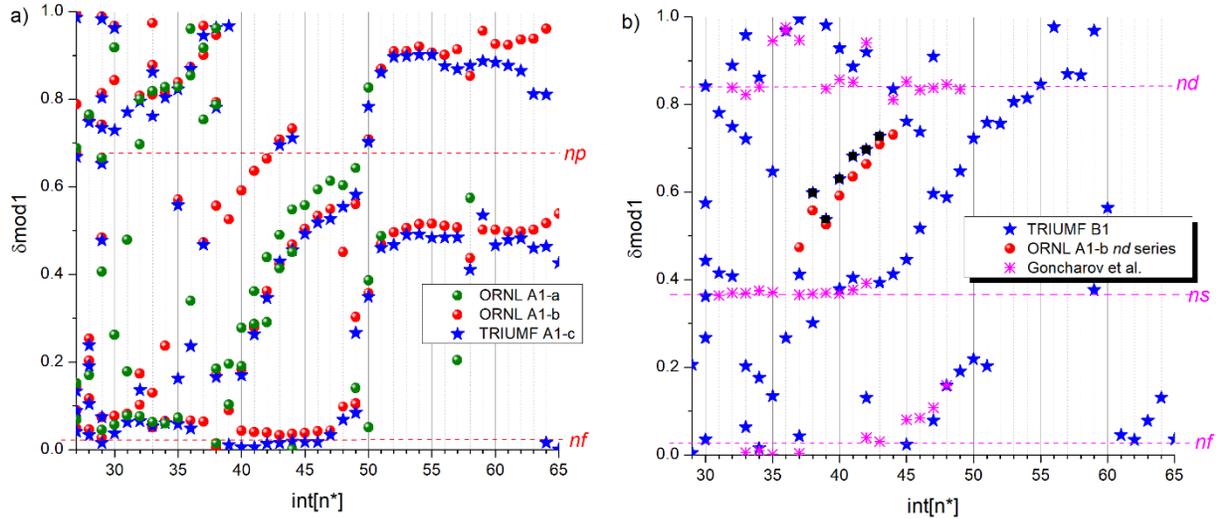

Fig. 4. The quantum defect $\delta$mod1 versus the effective quantum number $n^*$ for the observed states near and below the IP via four excitation schemes: A1-a, -b (ORNL), A1-c (TRIUMF), and B1 (TRIUMF). The values of $n^*$ and $\delta$ are obtained using Eqs. (2) and (3), based on the IP value 50868.735 cm$^{-1}$ determined in this work. a): Odd-parity states observed via three-step laser resonant excitations. b): Even-parity states observed via two-step laser resonance excitations, along with the forbidden $nd$ series from the A1-b three-step scheme. For comparison, the Rydberg states extracted from Goncharov's spectrum [25] are included in plot b) .

To assign these series properly, a Lu-Fano plot is made (Fig. 4a). The effective quantum number $n^*$ and the quantum defect $\delta$ are calculated using the Rydberg-Ritz formula, which describes how a Rydberg series $E_n$ converges to an ionization limit $E_{\text{limit}}$:

$$E_n = E_{\text{limit}} - \frac{R_M}{n^{*2}} = E_{\text{limit}} - \frac{R_M}{[n-\delta(n)]^2} \qquad (1)$$

$$n^* = \sqrt{\frac{R_M}{E_{\text{limit}} - E_n}} \qquad (2)$$

$$\delta(n) = n - n^* \qquad (3)$$

Here $n$ is the principal quantum number, $E_n$ is the corresponding level energy, $E_{\text{limit}}$ is the limit to which a series converges, and $R_M = 109737.0562$ cm$^{-1}$ is the reduced-mass Rydberg constant for $^{232}$Th.

Theoretical calculations [34] predict the quantum defects $\delta$ of the actinide atoms with $Z = 89 - 103$ to be approximately 5.2, 4.75, 3.8, and 2.0 for $ns$, $np$, $nd$, and $nf$ series, respectively. Experimental measurements of uranium ($Z = 92$) [35,36] showed $\delta$mod1 values of ~ 0.75 for the $np$ series and ~0.00 for the $nf$ series, respectively, consistent with the theoretical predictions. Goncharov et al. [25] reported $\delta'$ values of 0.639(6) and 0.984(31) for the $6d^27s$ $ns$ and $nd$ series of Th, respectively, where $\delta'$ is the fractional part of $n^*$, namely 1-$\delta$mod1. These $\delta'$ values correspond to $\delta$mod1 = 0.361 and 0.016, which deviate from the theoretical predictions of 0.2 for the $ns$ series and 0.8 for the $nd$ series of Th. Upon re-examining their spectra, we obtained $\delta$mod1 ≈ 0.36 for the $ns$ series and ≈0.85 for the $nd$ series, in much better agreement with the theoretical expectations. Based on the theoretical predictions and previous experimental data for U and Th, we anticipate $\delta$mod1 values for the $ns$, $np$, $nd$, and $nf$ series of Th to be approximately 0.36, 0.75, 0.85, and 0, respectively.

It is worth noting that we also found some Rydberg members in Goncharov's spectra that could belong to an additional series with $\delta mod1 \approx 0.07$, which is likely an $nf$ series (see Fig. 4b). Since the allowed Rydberg series in Goncharov's experiment should have even parity, the observation of the forbidden $nf$ series can be attributed to Stark effects caused by ambient stray electric fields [37] as Goncharov used the Mass-Analyzed Threshold Ionization (MATI) method, where a static electric field was applied to the photoexcitation region to separate ions from molecules. For high-lying Rydberg states, mix of the states with different $l$ by the Stark effect becomes prominent even in electric fields as low as 100 V/cm [38]. This is corroborated by our previous observations of the forbidden $np$ series of Be at $n = 30 - 45$ using RIS at the offline TRILIS stand [39] due to the stray electric field from the ion extraction field or the RF electric field inside IGLIS [40].

Perturbations from doubly excited states can cause significant deviations in the quantum defects of the Rydberg series from their expected value. As shown in Fig. 4, the observed Th Rydberg series exhibit substantial perturbations, particularly near $n^* = 38$ and $n^* = 50$, where large deviations in the quantum defects are observed. These perturbations also lead to pronounced variations in resonance ion signal intensity in the spectra around 50796 cm$^{-1}$ and 50825 cm$^{-1}$ (Fig. 3). Due to mixing with the perturber states, the excitation rates to the affected Rydberg states increase significantly, resulting in a noticeable change in resonance signal intensity. The irregular region can be highly localized to a single Rydberg state if the perturbation is weak; conversely, a strong perturbation can influence multiple Rydberg states. Such disruptions in the series regularity complicate the grouping and assignment of the observed Rydberg states.

Prior to the large perturbation near $n^* = 50$, two Rydberg series can be clearly distinguished. One has a nearly constant $\delta$mod1 ≈ 0.05, which matches with the theoretically predicted value for the $nf$ series and is in good agreement with the experimental values of the $nf$ series observed in uranium [36]. The other is strongly perturbed around $n^* = 38$, with its quantum defect changing significantly from $\delta$mod1 ≈ 0.75 at $n^* = 30$ to $\delta$mod1 ≈ 0.57 near $n^* = 50$, approaching an asymptotic value of $\delta$mod1 ≈ 0.68, which reasonably aligns with the theoretical value for the $np$ series. Hence, we can attribute the observed Rydberg states with $\delta$mod1 ≈ 0.05 in the region of $n^* = 30 - 48$ to the $nf_{5/2}$ or $nf_{7/2}$ series with $J = 3$, as they appear in the spectra excited from both $J = 2$ and $J = 4$ lower states. Similarly, the Rydberg states in the region of $n^* = 30 - 48$ with $\delta$mod1 deviating around 0.68, also excited from both $J = 2$ and $J = 4$ lower states, are assigned to the $np_{3/2}$ $J = 3$ series. One more Rydberg series was observed exclusively with scheme A1-a (excited from the $J = 2$ state) in the region of $n^* = 40 - 50$, exhibiting $\delta$mod1 values slightly higher than that of the $np_{3/2}$ $J = 3$ series. It is assigned to the series $np_{1/2}$ or $np_{3/2}$ with $J = 1$ or 2.

Additionally, via scheme A1-b, a distinct series is observed in the region $n^* = 37 - 44$ (Fig. 4), with $\delta$mod1 consistently higher than those of the $np$ series and gradually increasing from 0.47 to 0.73, approaching the expected value of 0.85 for the $nd$ series at higher $n$. The corresponding ion peaks of this series are very small, as shown in Fig. 5a, supporting its assignment to a forbidden $nd$ series of even parity. As discussed above, Stark mixing can enable otherwise forbidden transitions, such as the forbidden $np$ series observed in our previous resonance laser ionization spectroscopy [39], and the forbidden $nf$ series identified in Goncharov's MATI spectrum (Fig. 4b).

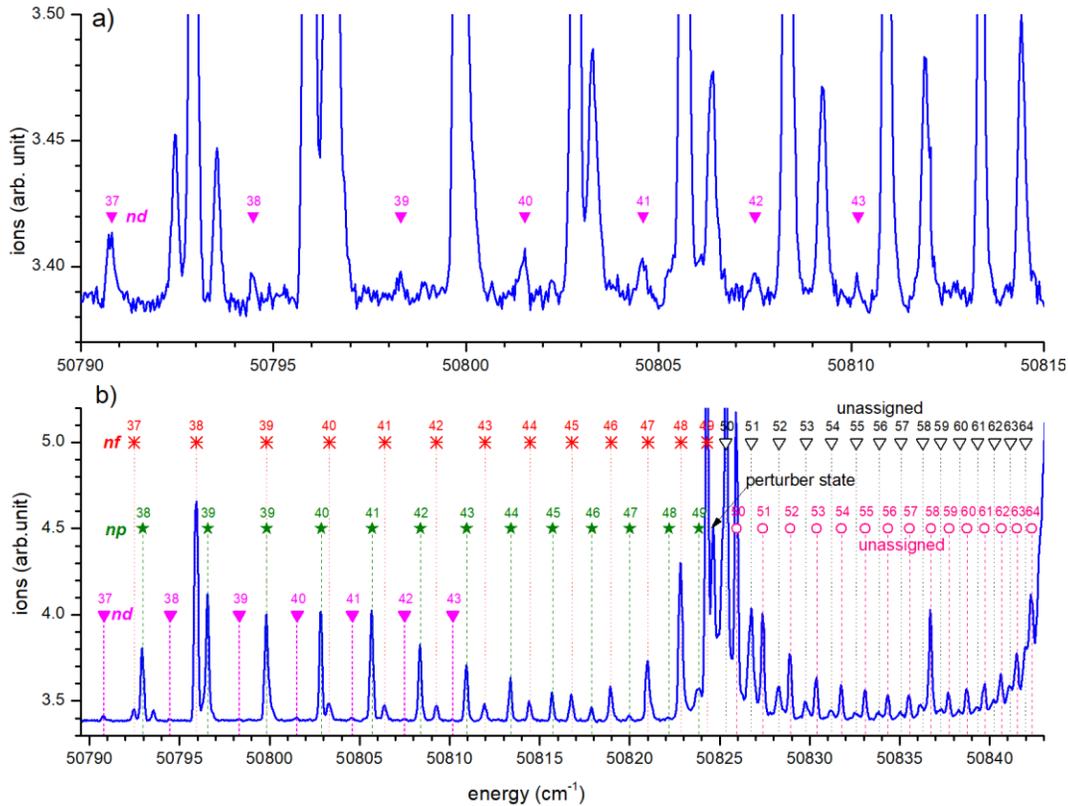

Fig. 5. Detailed spectra obtained using scheme A1-b at ORNL. The resonance peaks are labeled with the integer part of the effective quantum number, int($n^*$). a): Expanded spectrum with details about the tiny forbidden $nd$ resonances. b): Assigned Rydberg resonances before the strong perturbation at int($n^*$) $\approx 50$ ($np$, $nf$, and $nd$ series), and the unassigned resonance series after the perturbation.

After the perturbation near $n^* = 50$, two Rydberg series emerge with nearly constant $\delta$mod1 values that are different from those of the previously assigned $np$ and $nf$ series before $n^* = 50$. One series has $\delta$mod1 $\approx 0.5$ and the other has $\delta$mod1 $\approx 0.9$. The assignments for these two series cannot be unambiguously determined and are therefore labeled as unassigned in this paper. They cannot be simply interpreted as continuations of the $np$ and $nf$ series, since the observed quantum-defect shifts cannot be explained by perturbations, as perturbations merely cause temporary deviations of quantum defect near a perturber state and the quantum defect values return to normal once outside the perturbation region. For low-$l$ valence electrons, such as $p$-electrons, the quantum defect can vary with different $J$ (due to fine structure splitting), especially in heavy elements where such variations can be as large as 0.2. However, this is unlikely to be the case for the $nf$ series, due to its large $l$ value and very small interaction with the core. The $\delta$mod1 of $nf$ series with different $J$ are typically very close to 0. Therefore, the series with $\delta$mod1 $\approx$ 0.9 observed after $n^* = 50$ is unlikely to be the continuation of the earlier $nf$ series. Instead, considering the assigned $nd$ series in the region $n^* = 37 - 44$ and its trend toward the expected $\delta$mod1 value of 0.85 for an $nd$ series it is plausible that this series is also a forbidden $nd$ series. This series becomes observable due to higher sensitivity to Stark mixing in the high-$n$ Rydberg region. The other series observed beyond $n^* = 50$ has $\delta$mod1 $\approx$ 0.5, which lies between the expected $\delta$mod1 values of 0.36 and 0.68 for the $ns$ and $np$ series, respectively. It is difficult to determine whether this series corresponds to an $np$ series or a forbidden $ns$ series. Further experimental investigation or theoretical calculation is needed to clarify its assignment. The level energies and the effective quantum number $n^*$ of the assigned $6d^27s$ ($^4F_{3/2}$) $np$, $nd$, and $nf$ series are listed in Table 2. The two unassigned series with $n^* > 50$ are presented in Table 3. For the levels observed in multiple scans/measurements, the weighted average values are presented. The $n^*$ values in the tables are calculated using the newly determined IP of 50868.735 cm$^{-1}$, as derived in the next section.

Table 2. Energies of observed Rydberg states assigned to the $6d^27s$ ($^4F_{3/2}$) $np$ and $nf$ series converging to the IP of neutral Th. The listed energies and associated uncertainties are weighted averages of measurements obtained via schemes A1-a, A1-b, and A1-c. The relation between $n^*$ and the principal quantum number $n$ is $n \approx n^* + 2.05$ for the $nf$ series and $n \approx n^* + 4.68$ for the $np$ series. The states marked with "†" are strongly perturbed, with a mixture of the perturber states.

| $nf$, $J = 3, 4, 5$ | | $np_{3/2}$, $J = 3$ | | $np$, $J = 1, 2$ | |
|---|---|---|---|---|---|
| $E$ (cm$^{-1}$) | $n^*$ | $E$ (cm$^{-1}$) | $n^*$ | $E$ (cm$^{-1}$) | $n^*$ |
| 50746.26(5) | 29.93 | | | | |
| 50754.04(5) | 30.93 | 50756.22(15) | 31.23 | | |
| 50761.04(5) | 31.92 | 50762.87(5) | 32.20 | | |
| 50767.50(10) | 32.92 | 50769.11(5) | 33.19 | | |
| 50773.50(5) | 33.95 | 50774.82(5) | 34.18 | | |
| 50778.83(5) | 34.94 | 50780.00(5) | 35.17 | | |
| 50783.74(5) | 35.93 | 50784.68(5) | 36.13 | | |
| 50788.30(6) | 36.94 | 50788.97(5) | 37.09 | | |
| 50792.48(6) | 37.94 | 50792.93(5) | 38.05 | | |
| 50795.95(6) | 38.83 | 50796.58(5) | 39.00 | 50795.90(10) | 38.82 |
| 50799.88(6) | 39.92 | 50799.83(5) | 39.91 | 50799.47(10) | 39.80 |
| 50803.34(6) | 40.96 | 50802.87(5) | 40.82 | 50802.56(10) | 40.72 |
| 50806.42(6) | 41.97 | 50805.69(5) | 41.72 | 50805.44(10) | 41.64 |
| 50809.29(6) | 42.97 | 50808.38(6) | 42.64 | 50808.16(10) | 42.56 |
| 50811.98(6) | 43.97 | 50810.95(5) | 43.58 | 50810.77(10) | 43.51 |
| 50814.48(5) | 44.98 | 50813.41(5) | 44.54 | 50813.20(10) | 44.45 |
| 50816.80(6) | 45.97 | 50815.72(6) | 45.50 | 50815.59(10) | 45.44 |
| 50818.98(6) | 46.96 | 50817.91(6) | 46.47 | 50817.78(10) | 46.41 |
| 50821.02(6) | 47.96 | 50820.01(6) | 47.46 | 50819.87(10) | 47.39 |
| 50822.86(6) | 48.91 | 50822.15(6) | 48.54 | 50821.88(10) | 48.40 |
| 50824.31(6)† | 49.70 | 50823.83(6)† | 49.44 | 50823.69(10)† | 49.36 |

Table 3. Energies of unassigned Rydberg states observed with $n^* > 50$, converging to the IP of thorium. The listed energies and their uncertainties are weighted averages of measurements obtained via schemes A1-b and A1-c. The states marked with "†" are strongly perturbed, with a mixture of the perturber states.

| Series with $\delta$mod1 $\approx$ 0.9, $J = 3, 4, 5$ | | Series with $\delta$mod1 $\approx$ 0.5, $J = 3, 4, 5$ | |
|---|---|---|---|
| $E$ (cm$^{-1}$) | $n^*$ | $E$ (cm$^{-1}$) | $n^*$ |
| 50825.35(6)† | 50.29 | 50825.95(6)† | 50.65 |
| 50826.76(6)† | 51.13 | 50827.41(6) | 51.53 |
| 50828.29(6) | 52.09 | 50828.94(6) | 52.51 |
| 50829.80(6) | 53.09 | 50830.39(6) | 53.50 |
| 50831.21(6) | 54.08 | 50831.77(6) | 54.49 |
| 50832.58(6) | 55.09 | 50833.10(6) | 55.49 |
| 50833.87(6) | 56.11 | 50834.35(6) | 56.50 |
| 50835.07(6) | 57.09 | 50835.54(6) | 57.50 |
| 50836.28(6) | 58.14 | 50836.74(6) | 58.57 |
| 50837.27(6) | 59.06 | 50837.73(6) | 59.50 |
| 50838.34(6) | 60.08 | 50838.76(6) | 60.50 |
| 50839.33(6) | 61.09 | 50839.72(6) | 61.50 |
| 50840.26(6) | 62.08 | 50840.64(6) | 62.50 |
| 50841.16(6) | 63.08 | 50841.52(6) | 63.51 |
| 50842.00(6) | 64.06 | 50842.35(6) | 64.49 |
| | | 50843.21(15) | 65.57 |

At TRIUMF, a two-step scheme B1 was also employed to study the even-parity Rydberg states near the IP. The observed spectra are shown in Fig. 6. Excited from the $6d^2 7s7p\ ^3G^o_3$ intermediate state, the observed Rydberg series are expected to be even-parity $6d^2 7s\ (^4F_{3/2})\ ns$ or $nd$ series based on the electric-dipole-transition selection rule:

$$6d^2 7s7p\ ^3G^o_3 \rightarrow 6d^2 7s\ (^4F_{3/2})\ ns_{1/2},\ J = 2$$

$$nd_{3/2},\ J = 2,\ 3$$

$$nd_{5/2},\ J = 2,\ 3,\ 4$$

Fig. 4b is the Lu-Fano plot of the observed states via scheme B1, along with Goncharov's data (extracted from their spectrum) and the $nd$ states observed via scheme A1-b (Fig. 5a). The $nd$ series obtained from scheme A1-b agrees with the results from scheme B1, supporting the proposed configuration. Since being observed in both scheme A1-b (excited from a $J = 4$ state) and B1 (excited from a $J = 3$ state), it is assigned as either $nd_{3/2}\ J = 3$ or $nd_{5/2}\ J = 3, 4$. The other Rydberg series observed via scheme B1 (Fig. 4b and Fig. 6) are strongly perturbed. Owing to significant perturbations across a wide energy range and possible Stark mixing in the high-$n$ region, a definite assignment for this series is not currently possible. The Rydberg states with tentative assignments are listed in Table 3. The grouping of the Rydberg states and their tentative assignments were based on the regularity of resonance intensities in the spectrum (Fig. 6) and on the characteristic deviation-return behavior of the quantum defects about the expected values for the corresponding series (Fig. 4b). Further experimental investigations, as well as support through theoretical calculations, are needed to confirm/revise these assignments. The table lists only tentative assignments and is intended to make the level energy data available for future studies.

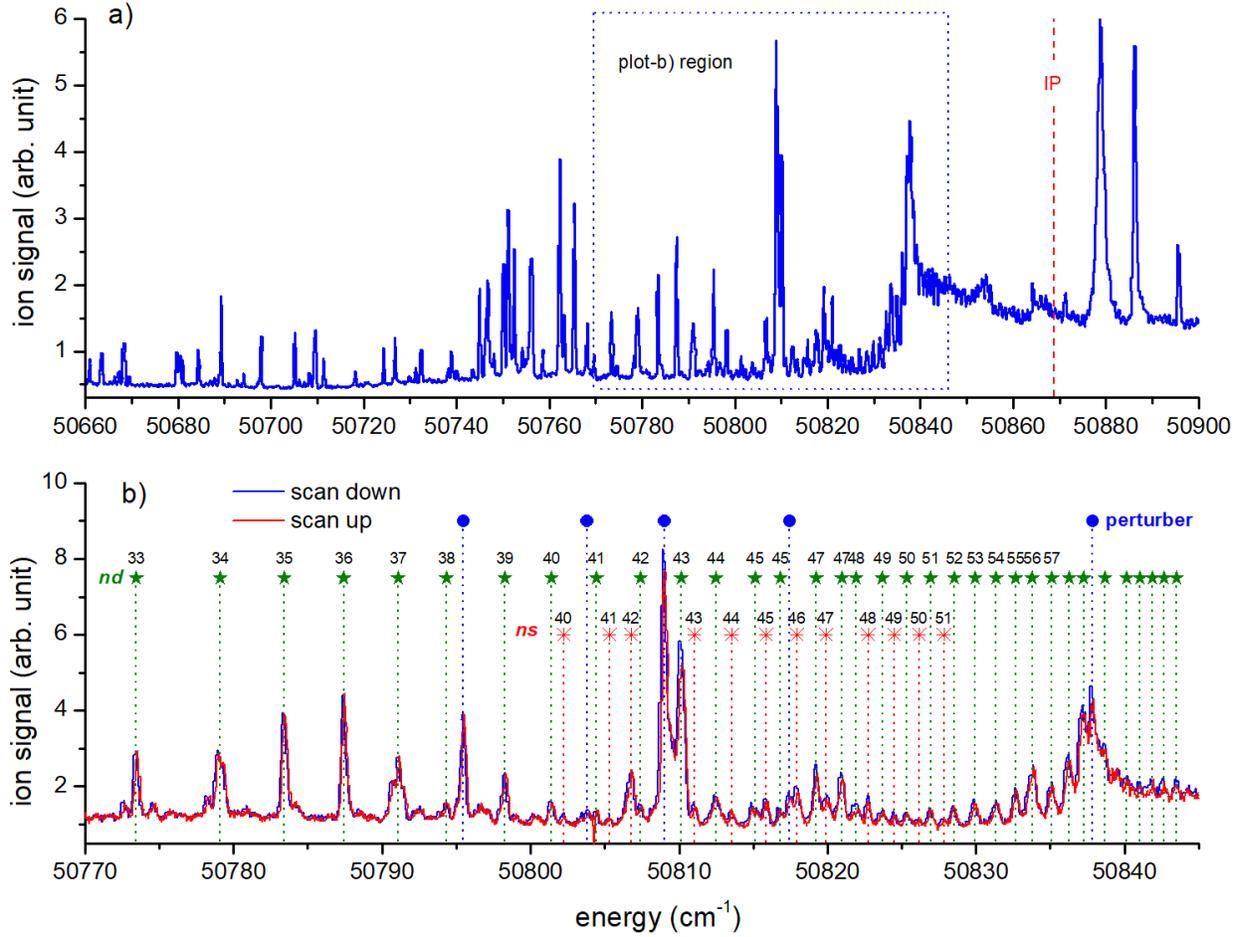

Fig. 6. The spectra obtained via the blue-blue scheme B1 at TRIUMF. a): Long-range spectrum including Rydberg states and three AI states. b): Short-range spectrum of the high-*n* Rydberg states, measured with slower scanning speeds and finer steps. The resonance peaks tentatively assigned as members of the *nd and ns* series are marked with int(*n\**).

Table 4. Even-parity Rydberg series observed via the two-step laser excitation scheme B1, tentatively assigned as $6d^2 7s$ ($^4F_{3/2}$) *nd* and *ns* series converging to the IP of Th. Unless explicitly marked, the uncertainty of the listed level energies is 0.15 cm⁻¹ which accounts for both statistical and systematic uncertainties. "*\**" marks the states observed too in the A1-b scheme (Fig. 5a), for which the listed energy and uncertainty are weighted average values.

| $nd_{3/2}$ $J = 3$ or $nd_{5/2}$ $J = 3, 4$ | | $ns_{1/2}$, $J = 2$ | |
|---|---|---|---|
| $E$ (cm⁻¹) | $n^*$ | $E$ (cm⁻¹) | $n^*$ |
| 50773.45 | 33.94 | | |
| 50779.08 | 34.98 | | |
| 50783.43 | 35.87 | | |
| 50787.41 | 36.73 | | |
| 50790.81(6)* | 37.53 | | |
| 50794.46(6)* | 38.44 | | |
| 50798.30(6)* | 39.47 | | |
| 50801.51(6)* | 40.40 | 50802.23 | 40.62 |
| 50804.58(6)* | 41.36 | 50805.31 | 41.60 |
| 50807.50(6)* | 42.33 | 50806.76 | 42.08 |
| 50810.18(5)* | 43.29 | 50811.03 | 43.61 |
| 50812.75(6)* | 44.27 | 50813.54 | 44.59 |
| 50815.11 | 45.24 | 50815.85 | 45.55 |
| 50816.82 | 45.98 | 50817.95 | 46.48 |
| 50819.25 | 47.09 | 50819.90 | 47.40 |
| 50820.95 | 47.92 | 50822.73 | 48.84 |
| 50821.91 | 48.41 | 50824.50 | 49.81 |
| 50823.68 | 49.35 | 50826.18 | 50.78 |
| 50825.32 | 50.28 | 50827.83 | 51.80 |
| 50826.94 | 51.24 | | |
| 50828.53 | 52.24 | | |
| 50829.95 | 53.19 | | |
| 50831.36 | 54.19 | | |
| 50832.66 | 55.15 | | |
| 50833.77 | 56.02 | | |
| 50835.11 | 57.13 | | |
| 50836.26 | 58.13 | | |
| 50837.24 | 59.03 | | |
| 50838.69 | 60.44 | | |
| 50840.15 | 61.95 | | |
| 50841.06 | 62.97 | | |
| 50841.88 | 63.92 | | |
| 50842.66 | 64.87 | | |
| 50843.52 | 65.96 | | |

## 3.2. Ionization potential of Th

The IP of neutral Th was measured to be 50867 cm$^{-1}$ with an uncertainty of 2 cm$^{-1}$ in 1994 [17], which remains the listed value in the NIST database [8]. More recently, Goncharov et al. [25] reported an IP of 50868.71(8) cm$^{-1}$ based on the convergence limit of the $ns$ Rydberg series. We have evaluated the IP value using the Rydberg series observed in this work by fitting the Rydberg-Ritz formula Eq. (1) to the members of the series with the Ritz expansion:

$$\delta(n) = n - n^* = \delta_0 + \frac{a}{(n-\delta_0)^2} + \cdots \qquad (4)$$

where $\delta_0$ is the asymptotic limiting value of the quantum defect $\delta$, $a$ is a constant, and the convergence limit $E_{limit}$ corresponds to the IP of Th. Higher order terms provide the correction for low-$n$ members where core penetration and polarization become significant. However, the Rydberg-Ritz formula is valid only for the series without perturbations. To accurately extract the IP value, only the unperturbed regions of the series can be used.

The Lu-Fano plot of the grouped levels in Tables 2 and 3 is shown in Fig. 7a. Three sets of the series members are selected for the determination of the IP: 1) the $nf$ series with $n^* = 29$ - 47; 2) the unassigned high-$n$ series with $\delta_t \mathrm{mod}1 \approx 0.45$ at $n^* = 51 - 64$; and 3) the unassigned high-$n$ series with $\delta_t \mathrm{mod}1 \approx 0.9$ at $n^* = 51 - 64$. Some members, i.e., $n^* = 38$ and 58, are clearly disturbed and are therefore excluded. Each data set is separately fitted to the Rydberg-Ritz formula, as shown in Fig. 7b. Since the quantum defect is nearly constant, $a$ in Eq. (4) is set to be 0 in the fitting. The convergence limits derived from the three data sets are summarized in Table 5. The relatively large uncertainties in the $\delta_t \mathrm{mod}1$ parameters for series 2) and 3) reflect the impact of the strong perturbation at int($n^*$) = 50 on the surrounding data points. The weighted average of the best-fit convergence limits gives IP = 50868.735(54) cm$^{-1}$. The uncertainty here is the $\sigma \sqrt{\chi_r^2}$, where $\sigma$ is the standard deviation of the three extracted IP values shown in Table 5, and $\chi_r$ is the reduced chi-squared of the three values scattering around the weighted average. Our result is in good agreement with the value reported by Goncharov et al. [25].

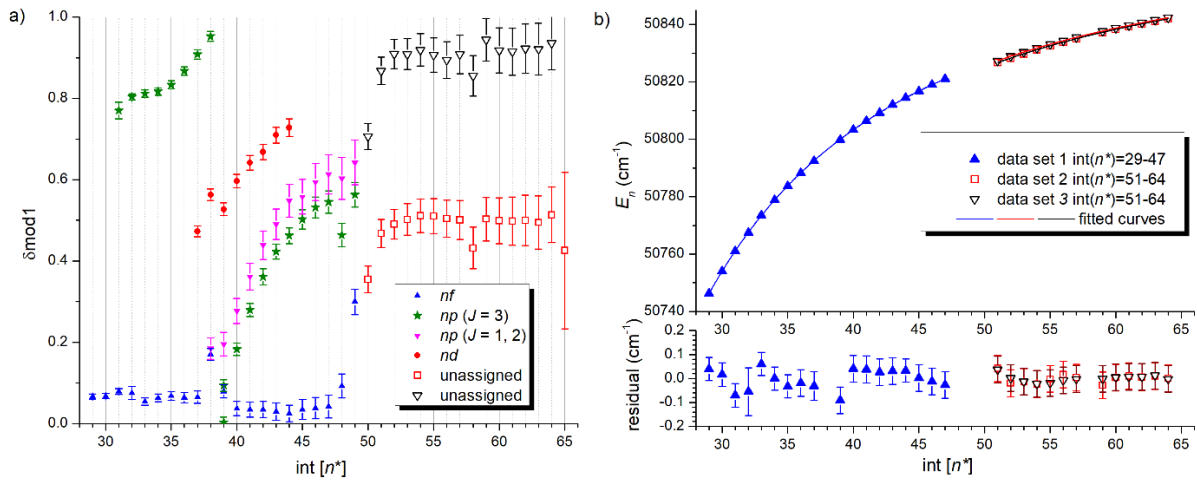

Fig. 7. a): Lu-Fano plot for the observed Rydberg series $6d^2 7s$ $(^4F_{3/2})$ $np$, $nf$, $nd$ series, and two unassigned series converging to the IP of Th. b): Rydberg-Ritz fits with the selected data sets and the residuals of the fits.

Table 5. IP values obtained by Rydberg-Ritz fits with the selected Rydberg series, compared with previous works.

| data set | $E_{\mathrm{limit}}$ (cm$^{-1}$) | $\delta_t \mathrm{mod}1$ |
|---|---|---|
| 1) $nf$ series, int($n^*$) = 29 – 47 (exclude 38) | 50868.848(25) | 0.085(5) |
| 2) unassigned, int($n^*$) = 51 – 64 (exclude 58) | 50868.702(21) | 0.470(17) |
| 3) unassigned, int($n^*$) = 51 – 64 (exclude 58) | 50868.663(26) | 0.850(21) |
| weighted average | 50868.735(54) | |
| NIST database and Trautmann [7,17] | 50867(2) | |
| Goncharov et al. [25] | 50868.71(8) | |

### 3.3 AI Rydberg spectra above the IP

AI Rydberg series were observed using the A2, A3 (blue-red-red) and B2, B3 (blue-blue) schemes within the energy range of 52300 – 52700 cm$^{-1}$. The corresponding spectra are shown in Figs. 8 and 9. For both three-step and two-step schemes, two series were identified, each converging to the 1$^{st}$ and 2$^{nd}$ metastable states of Th$^+$, respectively. These ionic metastable states are located at 1521.89632 cm$^{-1}$ and 1859.93843 cm$^{-1}$ above the IP, respectively. Based on the refined IP value of 50868.735 cm$^{-1}$ from this work, the two convergence limits are 52390.645 cm$^{-1}$ and 52728.687 cm$^{-1}$ with respect to the ground state of neutral Th. According to the NIST database [7], the configurations of the two metastable states are listed as $6d^2(^3F)7s\ ^4F\ J = 5/3$ and 3/2, respectively. However, as discussed above, we adopted $6d^2(^3F)7s\ ^4F_{3/2}$ as the configuration of the Th$^+$ ground state. To avoid confusion, we will use $6d\ 7s^2\ ^2D_{3/2}$ for the second metastable state. This second metastable state is greatly mixed, with 36% $6d\ 7s^2\ ^2D_{3/2}$ and 52% $6d^2(^3F)7s\ ^4F_{3/2}$.

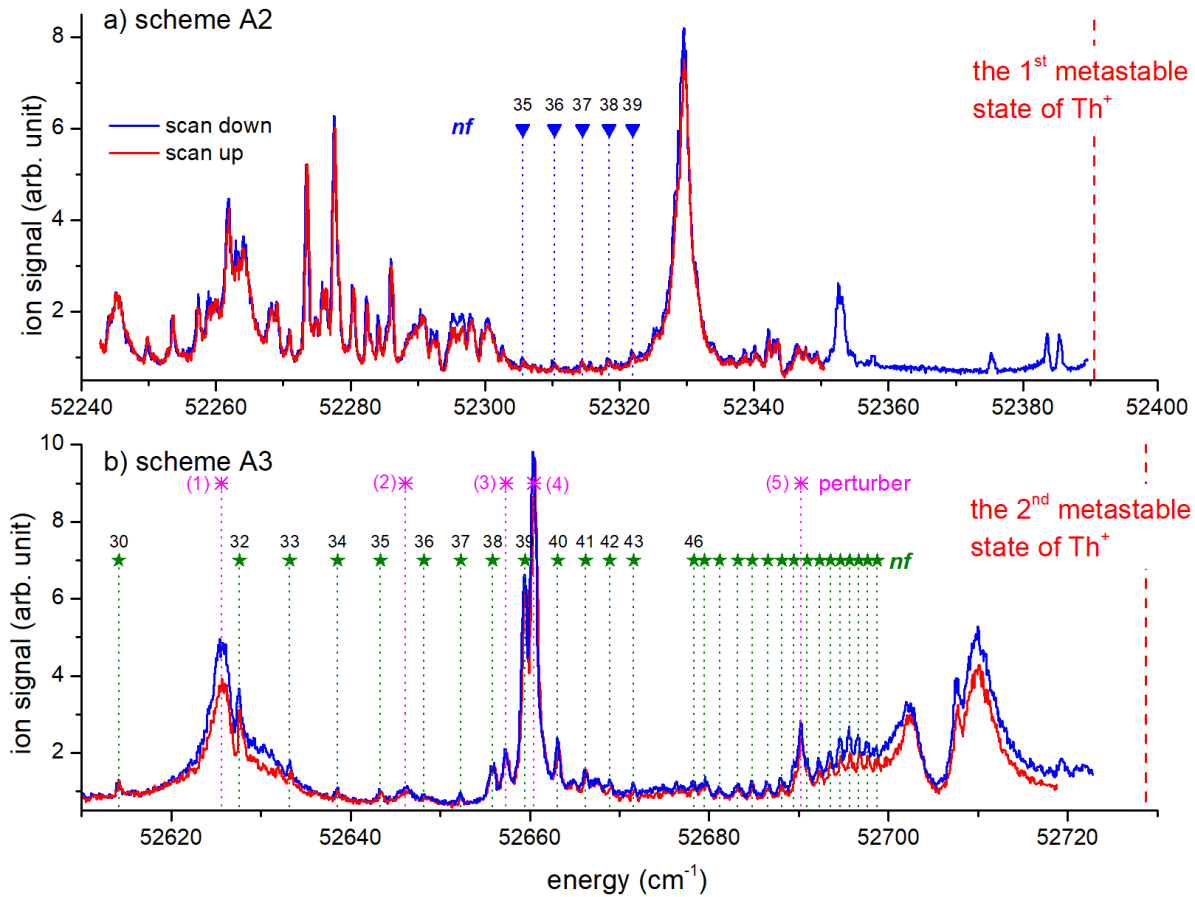

Fig. 8. AI spectra measured at TRIUMF via the blue-red-red schemes: a) scheme A2, b) scheme A3. Red traces correspond to scanning up and blue traces to scanning down. The scan laser power was 1.1 W. The first and second laser excitation steps were well saturated.

The energies of these AI Rydberg states have been extracted by fitting the resonances with Gaussian profile and are presented in Tables 6 and 7. The given values are the averages of the resonance energies obtained via two scanning directions, with an uncertainty of 0.15 cm$^{-1}$, which is the typical precision for Rydberg states measured via RIS at TRIUMF. In our previous analysis of systematic uncertainties in the measurement of AI Rydberg states [33], the standard deviation of the measured energies of 140 AI states was 0.4 cm⁻¹. However, in that work, the major sources of uncertainty were the wide profiles of the AI resonances, significant resonance overlapping, and complex resonance shape variations caused by interferences with nonresonant excitation to the continuum. In contrast, most of the AI states observed in this work exhibit sharp and symmetric resonances, without shape complications. This is attributed to the low excitation rate from the intermediate state $6d^2 7s 7p\ ^3 G^o_3$ to the underlying continuum states. There are also several wide resonances as marked with the larger uncertainty of 0.4 cm⁻¹ in Tables 6 and 7.

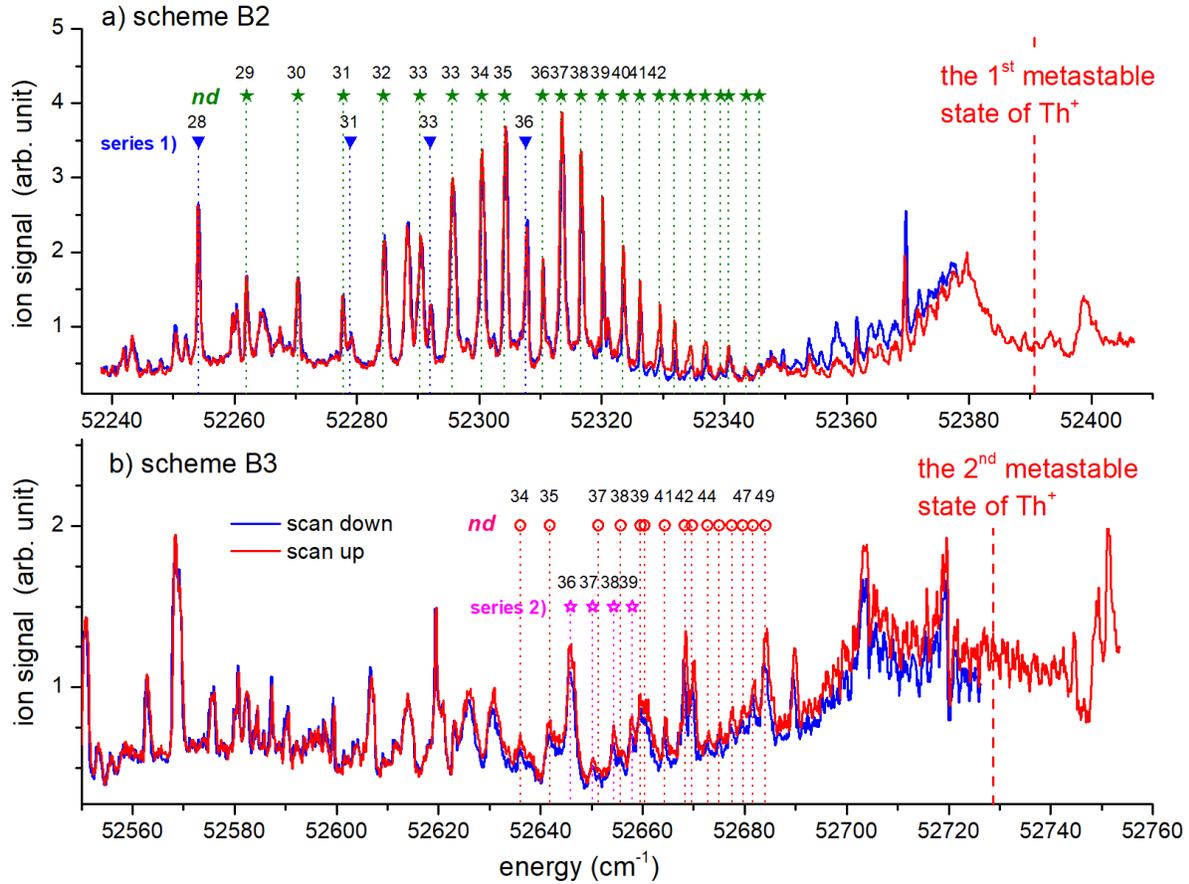

Fig. 9. AI spectra measured at TRIUMF via the blue-blue schemes: a) scheme B2, b) scheme B3. The red and blue traces represent scanning up and scanning down, respectively. The scan laser had a power of 450 mW. The first excitation step was well saturated. Two unassigned series are noted as series 1) and series 2).

The Lu-Fano plots of the AI series are shown in Fig. 10. Based on their $\delta$mod1 values, the two series observed using the blue-red-red scheme (Fig. 8) are assigned to be the $6d^2 7s (^4 F_{5/2}) nf$ and $6d7s^2 (^2 D_{3/2}) nf$ series of odd parity. The series observed with the blue-blue scheme (Fig. 9) are assigned to be the $6d^2 7s\ (^4 F_{5/2}) nd$ and $6d7s^2 (^2 D_{3/2}) nd$ series of even-parity. Two short-range series (marked as series 1 and 2 in Fig. 9 and Fig. 10b might be $np$ series via the Stark effect but are left unassigned due to insufficient supporting evidence. The perturber states marked in Fig. 8b and Fig. 10a not necessarily represent all existing perturber states. Obvious perturbations were observed only around the perturber states (2), (4), and (5).

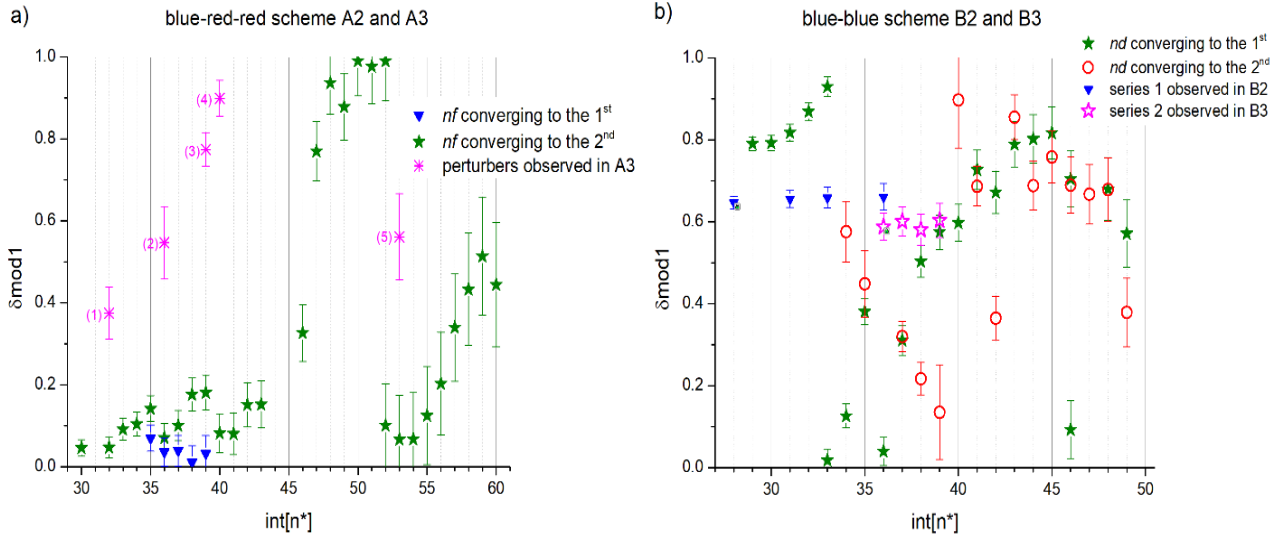

Fig. 10. Lu-Fano plots for the observed AI Rydberg series. a): observed series converging to the 1st metastable state of Th+ with blue-red-red schemes A2 and A3. The perturber states observed in A3 (marked in the same color in Fig. 8b) are also plotted. b): observed series converging to the 2nd metastable state of Th+ with blue-blue schemes B2 and B3.

Table 6. Odd-parity AI Rydberg series observed via blue-red-red schemes (A2 and A3) and assigned to the $6d^27s(^4F_{5/2})nf$ and $6d7s^2(^2D_{3/2})nf$ series converging to the 1st and 2nd metastable states of Th+, respectively. The uncertainty of the measured state energies is 0.15 cm⁻¹, except for those marked with "*", which have an uncertainty of 0.4 cm⁻¹.

| Series converging to the 1st metastable state of Th+ | | Series converging to the 2nd metastable state of Th+ | | | |
|---|---|---|---|---|---|
| $nf$, $J = 3, 4, 5$ | | $nf$, $J = 3, 4, 5$ | | perturber states | |
| $E$ (cm⁻¹) | $n^*$ | $E$ (cm⁻¹) | $n^*$ | $E$ (cm⁻¹) | $n^*$ |
| | | | 30.95 | | |
| | | 52627.61 | 32.95 | 52625.58* | 32.63 |
| | | 52633.23 | 33.91 | | |
| | | 52638.56 | 34.90 | | |
| 52305.63 | 35.93 | 52643.33 | 35.86 | | |
| 52310.32 | 36.96 | 52648.21 | 36.93 | 52646.09* | 36.45 |
| 52314.48 | 37.96 | 52652.27 | 37.90 | | |
| 52318.44 | 38.99 | 52655.87 | 38.82 | | |
| 52321.93 | 39.97 | 52659.46 | 39.82 | 52657.36 | 39.23 |
| | | 52663.13 | 40.92 | 52660.43 | 40.10 |
| | | 52666.22 | 41.92 | | |
| | | 52668.90 | 42.85 | | |
| | | 52671.60 | 43.85 | | |
| | | 52678.30 | 46.67 | | |

| | | | |
|---|---|---|---|
| 52679.48 | 47.23 | | |
| 52681.17 | 48.06 | | |
| 52683.19 | 49.12 | | |
| 52684.80 | 50.01 | | |
| 52686.52 | 51.02 | | |
| 52688.11 | 52.01 | | |
| 52689.46 | 52.90 | | |
| 52690.95 | 53.93 | 52690.25 | 53.44 |
| 52692.31 | 54.93 | | |
| 52693.52 | 55.88 | | |
| 52694.66 | 56.80 | | |
| 52695.67 | 57.66 | | |
| 52696.68 | 58.57 | | |
| 52697.66 | 59.49 | | |
| 52698.75 | 60.56 | | |

Table 7. Even-parity AI Rydberg series observed via blue-blue schemes (B2 and B3), assigned to the $6d^2 7s(^4F_{5/2})nd$ and $6d7s^2(^2D_{3/2})nd$ series, along with the unassigned Series 1 and 2, all converging to the 1st and 2nd metastable states of Th$^+$, respectively. Series 1 and 2 might be $np$ series observable due to the Stark effect. The uncertainty of the measured state energies is 0.15 cm$^{-1}$, except for those marked with "*", which have an uncertainty of 0.4 cm$^{-1}$.

| Series converging to the 1st metastable state of Th$^+$ | | | | Series converging to the 2nd metastable state of Th$^+$ | | | |
|---|---|---|---|---|---|---|---|
| $nd$, $J$ = 2, 3, 4 | | series 1, $J$ = 2, 3, 4 | | $nd$, $J$ = 2, 3, 4 | | series 2, $J$ = 2, 3, 4 | |
| $E$ (cm$^{-1}$) | $n*$ | $E$ (cm$^{-1}$) | $n*$ | $E$ (cm$^{-1}$) | $n*$ | $E$ (cm$^{-1}$) | $n*$ |
| | | 52254.13 | 28.35 | | | | |
| 52262.01 | 29.21 | | | | | | |
| 52270.37 | 30.21 | | | | | | |
| 52277.77 | 31.18 | | | | | | |
| 52284.34 | 32.13 | 52278.94 | 31.34 | | | | |
| 52290.29 | 33.07 | | | | | | |
| 52295.60 | 33.98 | 52291.92 | 33.34 | | | | |
| 52300.40 | 34.87 | | | 52636.07* | 34.42 | | |
| 52304.14 | 35.62 | | | 52641.85* | 35.55 | | |
| 52310.30 | 36.96 | 52307.53 | 36.34 | | | 52645.91 | 36.41 |
| 52313.38 | 37.69 | | | 52651.38 | 37.68 | 52650.22 | 37.40 |
| 52316.58 | 38.50 | | | 52655.72 | 38.78 | 52654.33 | 38.42 |
| 52320.03 | 39.43 | | | 52659.62* | 39.86 | 52657.97 | 39.40 |
| 52323.41 | 40.40 | | | 52660.44* | 40.10 | | |
| 52326.21 | 41.27 | | | 52664.38 | 41.31 | | |
| 52329.38 | 42.33 | | | 52668.30 | 42.64 | | |

| | | | |
|---|---|---|---|
| 52331.86 | 43.21 | 52669.72 | 43.14 |
| 52334.46 | 44.20 | 52672.79 | 44.31 |
| 52336.88 | 45.18 | 52675.06 | 45.24 |
| 52339.43 | 46.30 | 52677.51 | 46.31 |
| 52340.76 | 46.91 | 52679.69 | 47.33 |
| 52343.63 | 48.32 | 52681.68 | 48.32 |
| 52345.72 | 49.43 | 52684.11 | 49.62 |

## 4. Summary

Two-step and three-step resonance ionization spectroscopy have been employed at ORNL and TRIUMF to study Rydberg and AI Rydberg series of Th. The results from the two independent experimental setups show good agreement. Five Rydberg series were observed via four different intermediate states and are assigned as $nf$, $np$ (two), $nd$, and $ns$ series - all converging to the ionization potential of Th, i.e., the ground state $6d^2(^3F)7s\ ^4F_{3/2}$ of Th$^+$. The states at $n^* > 50$ are unassigned and need further investigation. From the analysis of these Rydberg series, the ionization potential of Th has been determined to be 50868.735(54) cm$^{-1}$, in excellent agreement with the value of 50868.71(8) cm$^{-1}$ reported by Goncharov et al. [25], and approximately 100 times more precise than the value of 50867(2) cm$^{-1}$ in NIST database [7,17]. The total uncertainty includes both systematic uncertainty from different experimental setups and statistical uncertainty using different Rydberg series. In addition, four AI Rydberg series converging to the 1$^{st}$ and 2$^{nd}$ metastable states of Th$^+$ were observed and assigned to $nf$ and $nd$ series. All observed Rydberg and AI Rydberg state energies are compiled for reference and future studies.

**Acknowledgments**

The experimental work is funded by TRIUMF which receives federal funding via a contribution agreement with the National Research Council of Canada and through Natural Sciences and Engineering Research Council of Canada (NSERC) Discovery Grant RGP-IN-2021-02993 to R. Li and SAP-IN-2017-00039 and RGP-IN-2024-05981 to J. Lassen. Y. Liu and D. W. Stracener acknowledge support from the U.S. Department of Energy, Office of Science, and Office of Nuclear Physics under contract number DE-AC05-00OR22725.